\documentclass[12pt]{article}

\usepackage{epsfig}

\newcommand{\be}{\begin{equation}}
\newcommand{\ee}{\end{equation}}
\newcommand{\sss}[1]{\mbox{\scriptsize #1}}

\newcommand{\I}{{\cal I}}

\newcommand{\OO}{{\cal O}}
\newcommand{\real}{{\cal\mbox{Re\,}}}

\renewcommand{\deg}{\mbox{$^\circ$}}

\newcommand{\GeV}{\unskip\,\mathrm{GeV}}
\newcommand{\MeV}{\unskip\,\mathrm{MeV}}

\newcommand{\pba}{\unskip\,\mathrm{pb}}

\begin{document}
\pagestyle{empty}

\begin{flushright}
DTP/00/44 \\                                                         
July 2000 
\end{flushright}

\vspace*{5mm}
\begin{center}
       {\bf IMPACT OF THE W BOSON DECAY WIDTH ON PHOTON \\ 
       BREMSSTRAHLUNG ACCOMPANYING W PAIR PRODUCTION} \\
  \vspace*{1cm} 
      {\bf A.P.~Chapovsky}$^{1}$,\ \  
      {\bf V.A.~Khoze}$^{1}$ \ \
      {\bf and} \ \ 
      {\bf W.J.~Stirling}$^{1,2}$\\ 
  \vspace{0.3cm}
     $^1$~{\it Department of Physics, University of 
     Durham, Durham DH1 3LE, England}\\
     $^2$~{\it Department of Mathematical Sciences, University of 
     Durham, Durham DH1 3LE, England}\\
  \vspace{2cm}
      {\bf ABSTRACT} \\ 
\end{center}
\vspace*{5mm}
\noindent 
      The bremsstrahlung spectrum of photons accompanying $W$ pair production
      depends on the decay width of the unstable $W$ bosons. The dependence
      arises from the 
      interference between emission at different stages of the production and
      decay process.
      We present a quantitative discussion of this effect, and  consider the
      implications for measurements at LEP2 and LC energies.
\vspace*{2cm}\\ 

\vfill
\noindent 
\rule[.1in]{16.5cm}{.002in}

\noindent
emails: A.P.~Chapovsky@durham.ac.uk, 
V.A.Khoze@durham.ac.uk, W.J.Stirling@durham.ac.uk      

%
\newpage
\setcounter{page}{1}
\pagestyle{plain}


\section{Introduction}

Operating a future electron--positron linear collider (LC) in the LEP2 energy region but with 
much higher luminosity would allow a variety of precision tests of the electroweak sector to
be performed, see for example Ref.~\cite{LC-review}.
For instance, a 100~fb$^{-1}$ scan of the $W^+W^-$ threshold with longitudinally polarized electrons
and positrons would offer an opportunity to measure the $W$-boson mass with an error of $6\MeV$,
\cite{wilson}, with negligible uncertainty from QCD interconnection effects, see for example 
Ref.~\cite{sjos-khoze}. 
Such a threshold scan could also potentially provide a precise measurement of the $W$ decay
width, $\Gamma_{W}$.

At present, the most precise determination of $\Gamma_W$ comes from the 
indirect measurement at the Tevatron using the ratio of dilepton $Z$ and $W$ events \cite{tevatron-width}:
\begin{equation}
\Gamma_W({\rm CDF}+{\rm D0},\ \mbox{indirect})\; = \; 2.171 \; \pm \; 0.027\; (\mbox{stat.}) 
\; \pm \; 0.056   \;(\mbox{sys.})\ \GeV
\end{equation}
i.e.  with an overall precision
of approximately $60\MeV$. However it is very important to perform a {\it direct} 
determination of this key parameter of the Standard Model as well. 
The accuracy of the direct measurement
of the $W$ width is still not very high. Recently the CDF collaboration at the Tevatron have reported the 
value  \cite{CDF-width}:
\begin{equation}
\Gamma_W({\rm CDF},\ \mbox{direct})\; = \; 2.055 \; \pm \; 0.100\; (\mbox{stat.}) 
\; \pm \; 0.075   \;(\mbox{sys.})\ \GeV
\end{equation}
{}from measurements of the transverse mass spectrum in  leptonic $W$ decays.
Finally, the LEP experiments have made a preliminary measurement of the $W$ width from 
the line shape in $W$ pair production \cite{lep-width}:
\begin{equation}
\Gamma_W({\rm LEP2},\ \mbox{direct})\; = \; 2.19 \; \pm \; 0.15\; (\mbox{stat.}+\mbox{sys.}) 
\ \GeV
\end{equation}
Another related method, not yet exploited, would be to perform 
a precision scan of the $WW$ cross section in the threshold region, for example
at a future linear collider.

In this paper we focus on another method which is based on
previous studies in Ref.~\cite{dkos}.
This exploits the high sensitivity of soft photon radiation in pair production of  
$W$ bosons in $e^+e^-$ collisions 
to the $W$ width. Since the event rate is $\OO(\alpha)$ relative to the 
total $WW$ cross section, this method is potentially limited by statistics.
However it has the advantage of  avoiding problems such  as the effect of
beamsstrahlung or beam energy spread on line shapes.
Whether it will ultimately  be statistically competitive with a threshold scan in
the  precise determination of $\Gamma_W$ will require a dedicated analysis, 
which is beyond the scope of our studies here. However, our results do suggest 
that such further investigation would certainly be worthwhile.

The paper is organized as follows. In the next section we discuss the overall
structure of the $e^+e^-\to W^+W^-\gamma$ differential cross section,
emphasizing the various energy regimes for the photon radiation. We present analytic
results for the cross section in the soft-photon limit, which allows us 
to identify the factorizable and non-factorizable contributions. In Section~3 we discuss
the various ways of enhancing the non-factorizable contributions, which contain the 
bulk of the $\Gamma_W$ dependence, by imposing angular cuts on the final-state
particles. We illustrate our results by numerical calculations, considering the 
various possible leptonic and hadronic decay channels.
We also mention briefly the analogous results for the $\gamma\gamma\to W^+W^-\gamma$
process. Section~4 contains our conclusions.

\section{Bremsstrahlung radiation pattern in \boldmath $W^{+}W^{-}$ production}

The general formalism for calculating the soft radiation pattern 
in processes involving the  production and decay of unstable particles 
can be found in Refs.~\cite{kos,dkos,ww-dpa,ww-denner,nf}.
It is well known that heavy unstable particles such as the $W$ boson can radiate  
before and after their decay.
The relative intensity of the two contributions, and consequently the overall structure 
of the radiation pattern, depends sensitively on the relative size of the emission time-scale 
and  the particle lifetime, see e.g. \cite{dkos,kos}.
In particular, in the second reference in \cite{dkos} one can find a semi-classical explanation of 
how the radiation pattern allows the relative distance between the $W$-boson decay
vertices to be probed.\footnote{This phenomenon resembles an old idea 
\cite{eisberg-yennie} to use  soft-photon radiation
for measuring the time delay in nuclear reactions.}

We begin by recalling the main properties of the differential distribution for the radiation 
of a soft photon with momentum $k^{\mu}$ in the process
\be
\label{ee->ww->4f}
        e^{+}(q_1)e^{-}(q_2)\;	 \to\;	 W^{+}(p_1)W^{-}(p_2)\;	 \to\;	 
        4\ \mbox{fermions}(p_3,p_3',p_4,p_4') 
     \;	\biggl[+ \gamma(k)\biggr]\; .
\ee
It is well known that when unstable particles are produced 
one is forced to perform a Dyson resummation, which
leads to the regularization of the singularities in the propagators 
$1/(M_{1,2}^2-M_W^2)\to 1/D_{1,2}$, where  $D_{1,2}=M_{1,2}^2-M_W^2+i M_W\Gamma_W$, and $M_{1,2}$
are the invariant masses of the $W$ bosons.  
However such resummation leads in general to the breaking of gauge invariance 
through higher-order contributions picked up by Dyson resummation. Thus
this standard perturbative approach does not produce an acceptable  gauge independent answer.
The problem can be avoided by working in the so-called `pole-scheme' \cite{ww-pole-scheme}. 
The physical picture behind the pole-scheme is as follows.
Any process involving unstable particles can be viewed as a consequence of 
several subprocesses:
{\it production}, which is a hard process with a short time-scale $\OO(1/M_W)$;
{\it propagation} over a typically larger time $\OO(1/\Gamma_W)$;
and {\it decay}, which is again a hard process with a time scale $\OO(1/M_W)$. 
Technically, in the perturbative expansion gauge invariance is guaranteed only order by order.
Dyson resummation mixes different orders of the perturbative expansion, and thus breaks gauge
invariance. In order to restore it one has to re-expand the amplitudes again in some 
physical parameter other than the coupling constant, in a way that does not produce singularities. 
An appropriate  small parameter is $\Gamma_W/M_W$. It is constructed as a ratio 
of two physical scales: the scale of production and decay, $M_W$,
and the scale of propagation, $\Gamma_W$.
It should be noted  that this is a somewhat simplified picture, since sometimes there are 
additional small parameters present in the problem 
(like the relative velocity, $\beta$, close to threshold, or $M_W^2/s$ 
at ultra-relativistic energies, etc.). 
Then the above mentioned estimates may change, but the arguments remain similar. 

{}From the above considerations one can estimate the accuracy of calculations performed in 
the pole-scheme. When examining the process (\ref{ee->ww->4f}) one distinguishes three energy 
domains classified by the  distance in energy from threshold, 
$\Delta E=\sqrt{s}-2M_W$, compared to the relevant scales of the process, $\Gamma_W$ and $M_W$:
\begin{itemize} 
 	\item Relativistic region, $\Delta E\sim M_{W}$, where the accuracy is $\OO(\Gamma_W/M_W)$.
        \item `Far-from-threshold' region, $\Gamma_{W}\ll \Delta E\ll M_{W}$,
                where the accuracy is $\OO(\Gamma_W/\Delta E)$.
        \item Threshold region, $\Delta E\sim \Gamma_W$, where the accuracy is $\OO(1)$, 
        and the pole-scheme expansion breaks down.
\end{itemize}
The pole-scheme approach to processes involving unstable particles has been used to calculate
the full $\OO(\alpha)$ correction to the pair production of $W$ bosons in $e^+e^-$ 
collisions  \cite{ww-dpa, ww-denner}.
In this paper we use the results of Refs.~\cite{dkos,ww-dpa} as a basis for the calculations.
Because of the way the pole-scheme is constructed, one can classify all the radiative corrections 
into two types: {\it factorizable}, which act inside 
separate hard subprocesses (production and decay);
and {\it non-factorizable}, which interconnect various hard subprocesses. Here we will concentrate 
on the real photon radiation from $W$-pair production in the LEP2/LC energy region $170-500\GeV$. 
Again, there are three regimes for photon radiation.
\begin{itemize} 
 	\item Hard photon radiation, $\omega\sim M_W$, when the photon wavelength is 
 	of the same order as the hard process time-scale. The photon can be assigned to 
 	one of the hard subprocesses. Alternatively one can say that the photons radiated 
 	from different stages of the process do not interfere with each other.
        The radiation is exclusively factorizable.
        \item Soft radiation, $\omega\ll \Gamma_W$, when the photon wavelength is much larger 
		than the propagation distance. 
		In this case the photons cannot distinguish the details of the process and are
		radiated coherently from all stages of the process.
		Both factorizable and non-factorizable contributions are important.
        \item Semi-soft radiation,  $\omega\sim \Gamma_W$, when the photon wavelength is of the same
        order as the distance between the $W$ decay vertices. 
		In this case both factorizable and non-factorizable contributions are important.
		However photons are not radiated coherently from all stages of the process.
\end{itemize}
{}From this classification one can see that at $\omega\sim\Gamma_W$ there is a transition from 
a regime in which various subprocesses do not interfere with each other
to a regime in which the  photon does not distinguish details of the process.
This is this transition that is of interest to us in this paper, since it is where the photon
spectrum has maximum sensitivity to $\Gamma_W$. 

Note that when $\omega\sim\Gamma_W$ the photon is soft with respect to the hard scale of the 
process, $\omega\ll M_W$, but not with respect to the soft scale of the process, $\Gamma_W$.
As a consequence the cross section has  certain factorization properties.
The hard ($M_W$-scale) part of the amplitude factorizes just as in the conventional soft-photon approximation, but the soft ($\Gamma_W$-scale) part does not always factorize in the conventional way. 
This is why we call photons with 
 energy $\omega\sim\Gamma_W$ `semi-soft', rather than simply `soft'.    
Making use of the factorization properties, the complete radiation distribution 
in this semi-soft regime 
can be written as an interference of semi-soft currents with the hard parts of the amplitudes
in the following way:
\be
\label{mtrx}
        d\sigma
        =
        -\, d\sigma_{\sss{Born}}
        \frac{d\vec{k}}{(2\pi)^{3}2k_{0}}
        \Biggl[
        2\, \real\biggl(\I_{0}^{}\cdot\I_{+}^{*}+\I_{0}^{}\cdot\I_{-}^{*}
              +\I_{+}^{}\cdot\I_{-}^{*}\biggr)
         +\I_{0}^{}\cdot\I_{0}^{*}+\I_{+}^{}\cdot\I_{+}^{*}+\I_{-}^{}\cdot\I_{-}^{*}
         \Biggr]\; .
\ee
Here $d\sigma_{\sss{Born}}$ is the  Born cross section in the pole approximation. 
The currents $\I_{0}$ and $\I_{\pm}$ correspond to the radiation from the production
and decay stages respectively. 
The first three terms are non-factorizable contributions, 
consisting of final--final, $\I_{+}\I_{-}^{*}$, 
and initial--final, $\I_{0}\I_{+}^{*}+\I_{0}\I_{-}^{*}$, state interferences.
The last three terms are the factorizable contributions 
corresponding to  the production and decay parts.
The gauge-invariant semi-soft currents $\I_{0}$ and $\I_{\pm}$ are given by
\begin{eqnarray}
\label{currents}
 \I_{0}^{\mu}
 &=& +\,e\Biggl[ \frac{p_{1}^{\mu}}{k\cdot p_{1}}-\frac{p_{2}^{\mu}}{k\cdot p_{2}}
              -\frac{q_{1}^{\mu}}{k\cdot q_{1}} + \frac{q_{2}^{\mu}}{k\cdot q_{2}}
      \Biggr]\; ,
      \nonumber \\
 \I_{+}^{\mu} 
 &=& -\,e\Biggl[ \frac{p_{1}^{\mu}}{k\cdot p_{1}}
               +Q_{f_3}\frac{p_{3}^{\mu}}{k\cdot p_{3}}
               -Q_{f_3'}\frac{{p_{3}'}^{\mu}}{k\cdot p_{3}'}
       \Biggr]\frac{D_{1}}{D_{1}+2k\cdot p_{1}}\; ,
      \nonumber \\ 
 \I_{-}^{\mu} 
 &=& +\,e\Biggl[ \frac{p_{2}^{\mu}}{k\cdot p_{2}}
               + Q_{f_4}\frac{p_{4}^{\mu}}{k\cdot p_{4}}
               - Q_{f_4'}\frac{{p_{4}'}^{\mu}}{k\cdot p_{4}'}
       \Biggr]\frac{D_{2}}{D_{2}+2k\cdot p_{2}}\; .
\end{eqnarray}
The factors $Q_f, Q_{f'}$ are the electric charges of the final-state fermions,
with $Q_f - Q_{f'} =  - 1$. 
Recall that the integration over the invariant masses of the unstable particles eliminates
the pre-factors $D_{1,2}/(D_{1,2}+2k\cdot p_{1,2}) $ in the factorizable terms. 
In this case the semi-soft currents become the usual soft-photon ones, and
factorization takes place with respect to both scales of the process, hence
the name `factorizable'.
In the non-factorizable contributions, however, non-trivial pre-factors survive,
and complete factorization with respect to both scales does not take place.
An important consequence of this non-factorization is that for hard photons the 
non-factorizable contribution is suppressed because of the photon energy dependence
in the pre-factors, see also Refs.~\cite{dkos,kos,fkm}.
Thus non-factorizable contributions are important only for soft and semi-soft photons.
\begin{figure}[t]
  \unitlength 1cm
  \begin{center}
 \begin{picture}(8,7)
  \put(-0.7,6){\makebox[0pt][c]{\boldmath\small $\frac{1}{\sigma_{\sss{Born}}}\omega\frac{d\sigma}{d\omega}$}}
  \put(-0.2,5){\makebox[0pt][c]{\boldmath\small [\%]}}
  \put(6.5,-0.3){\makebox[0pt][c]{\boldmath\small$\omega, GeV$}}
  \put(0.2,0){\epsfig{file=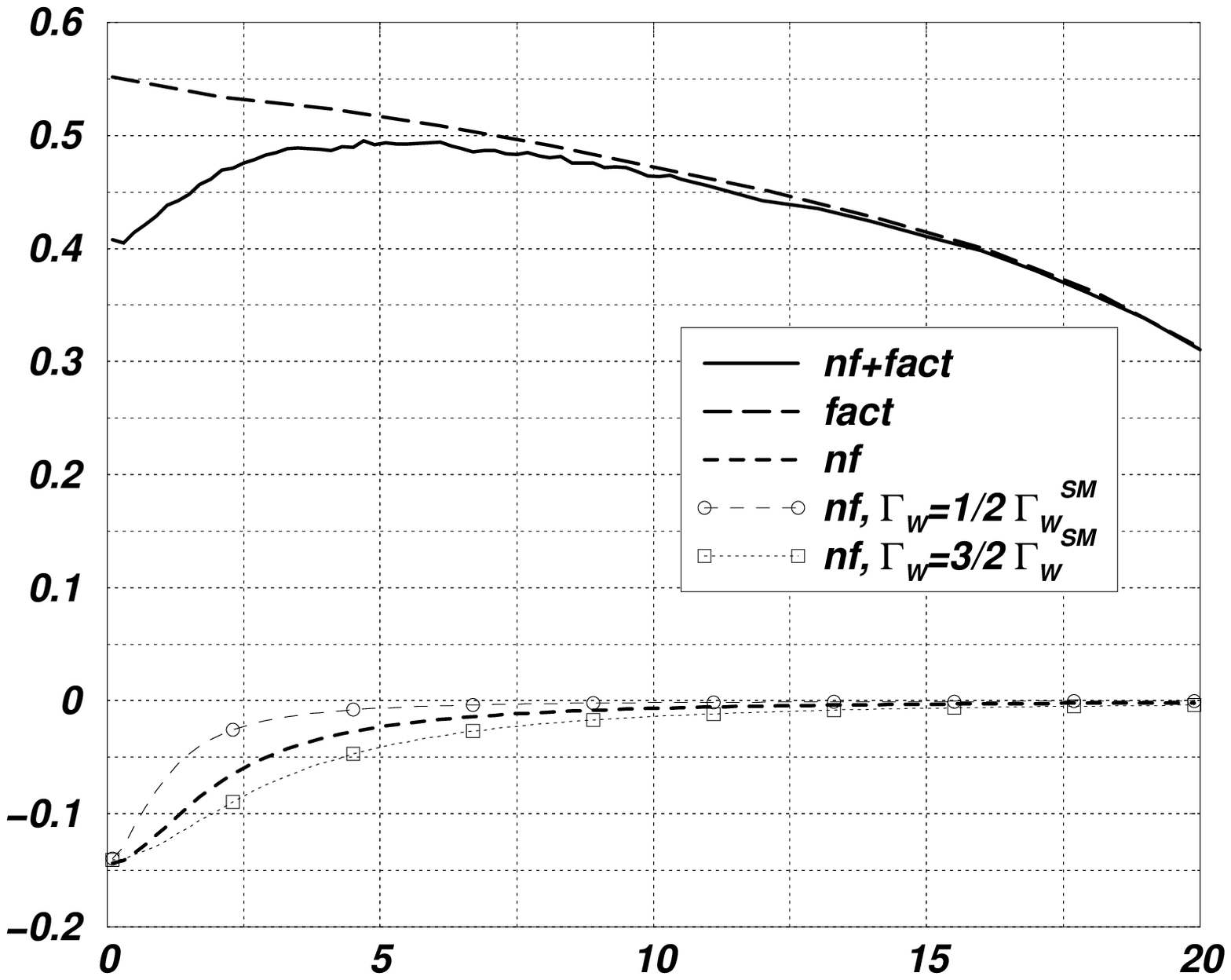,height=7cm,width=7cm,angle=0}}
  \end{picture}
  \end{center}
  \caption[]{The photon energy distribution, 
  $\frac{1}{\sigma_{\sss{Born}}}\omega\frac{d\sigma}{d\omega}$, 
		normalized to the Born cross-section, $\sigma_{\sss{Born}}\approx0.22\pba$,
		as a function of the photon energy, $\omega$, 
		in the semi-soft regime. 
		The final state is $(\mu^+\nu_\mu)(\tau^-\bar{\nu}_\tau)$ and the $e^+e^-$ 
		CMS energy is $\sqrt{s}=184\GeV$. The photon isolation cut, which restricts the
		photon to be separated by 
		at least $50\deg$ from all the charged fermions, is applied.
		Also shown is the non-factorizable contribution calculated for
		various values of the width, $\Gamma_W=(1/2;1;3/2) \Gamma_W^{SM}$,
		illustrating strong width dependence.}
\label{fig:spectrum}
\end{figure}

The qualitative picture described above is illustrated quantitatively
in Fig.~\ref{fig:spectrum}, which shows the photon energy spectrum,
$1/\sigma_{\sss{Born}}\ \omega \, d\sigma/d\omega$, is shown as a function of the
photon energy, $\omega$, in the semi-soft region.%
\footnote{Here and below we use the results and parameter values 
of Ref.~\cite{ww-dpa} for numerical calculations. In particular we use the Standard Model
$W$-boson width $\Gamma_W=2.082\GeV$.}
In this example the $e^+e^-$ CMS energy is $\sqrt{s}=184\GeV$ 
and a purely leptonic  $(\mu^+\nu_\mu)(\tau^-\bar{\nu}_\tau)$ final state
is chosen. 
A photon `isolation' cut is  also applied. This requires that in the CMS frame
the direction
of the radiated photon is separated by at least $50\deg$  from  
the directions of all the experimentally observed
charged particles (i.e. the initial-state $e^{\pm}$ and the final-state $\mu$ and $\tau$
leptons). By imposing these
`no-flight' zones around the charged particles we avoid the quasi-collinear-singularities
inherent in the currents in (\ref{currents}).~\footnote{In practice, the collinear
singularities are regulated by non-zero fermion masses, see below.}

We can see from Fig.~\ref{fig:spectrum} that the (negative) 
 non-factorizable contribution to the cross section is 
indeed suppressed for hard photons, with the 
damping occurring in the semi-soft regime of the photon energy, $\omega\sim\Gamma_W\approx2\GeV$.
In the same photon energy region  the dependence of the 
factorizable contribution on $\omega$ is practically flat.
This leads to a peaking  behaviour of the complete spectrum in the 
semi-soft region, $\omega\sim\Gamma_W$.  Thus by comparing the measured 
photon bremsstrahlung distribution in this region with the theoretical prediction
regarded as a function of $\Gamma_W$, 
one can in principle  
determine the $W$-boson width, as advocated in Ref.~\cite{dkos}. 

In order to gain some quantitative insight on how the method may work in practice,
two issues are important:
\begin{itemize} 
 	\item How pronounced is the shape of the relevant part of the
 	photon spectrum? 
		In other words, how large is the interesting (strongly $\Gamma_W$
		dependent, see Fig.~\ref{fig:spectrum}) 
		non-factorizable contribution with 
		respect to the factorizable contribution?
		The relevant parameter here is 
		\be
			\alpha(\sqrt{s}, \mbox{cuts}) = 
			\Biggl.
			\frac{(d\sigma_{\sss{nf}}/d\omega)}
			       {(d\sigma_{\sss{fact}}/d\omega)}
			\Biggr|_{\omega\to0}\; ,
		\ee 
		which  depends on the system of cuts chosen and the collider energy, $\sqrt{s}$.
        \item How large is the statistics for a particular choice of cuts? 
		The relevant parameter is the corresponding  Born cross section
		restricted by a particular system of cuts,	
		$d\sigma_{\sss{Born}}(\sqrt{s}, \mbox{cut})$.
\end{itemize}
\begin{figure}[t]
  \unitlength 1cm
  \begin{center}
 \begin{picture}(8,7)
  \put(-0.2,6){\makebox[0pt][c]{\boldmath\small $\alpha$}}
  \put(6.5,-0.3){\makebox[0pt][c]{\boldmath\small$\sqrt{s}, GeV$}}
  \put(0.2,0){\epsfig{file=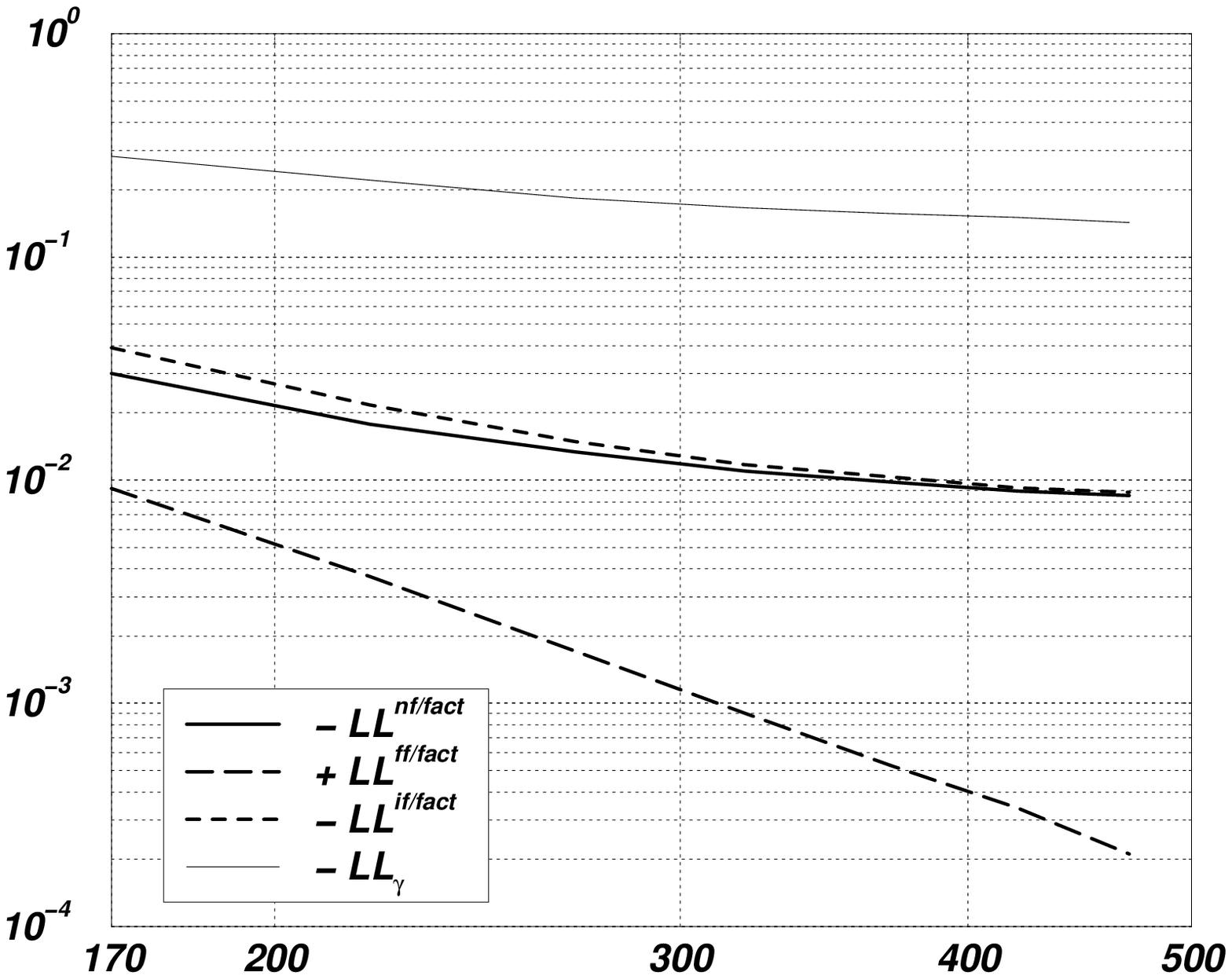,height=7cm,width=7cm,angle=0}}
  \end{picture}
  \end{center}
  \caption[]{The absolute value of the ratio of non-factorizable and 
  factorizable contributions to the cross section, $\alpha$, 
		for the leptonic $\mu\tau$ final state,
		as a function of the CMS energy with ($LL_\gamma$)
		 and without ($LL^{\sss{nf/fact}}$) photon isolation
		cuts ($ > 50\deg$ from all charged particles).
		The sign of the ratio is show in the legend of the plot.
		In the case of no cuts, the 
		initial-final, $LL^{\sss{if/fact}}$,  and final-final, $LL^{\sss{ff/fact}}$, 
		interferences are shown separately.}
\label{fig:nocuts}
\end{figure}
To illustrate how the  ratio of non-factorizable to 
factorizable radiation, $\alpha(\mbox{no cuts},\sqrt{s})$,
is influenced by the photon  isolation  cuts, we show 
in Fig.~\ref{fig:nocuts} the ratio as a function of the CMS energy, $\sqrt{s}$,
with and without cuts. 
In order to use a logarithmic scale we plot the absolute values of the ratios, and
indicate their sign in the legend of the plot.
Here and in what follows we label all quantities by two letters, which specify the decay 
channel of each of the $W$'s,  $L$ for leptonic and $H$ for hadronic,
and a subscript, which specifies the system of cuts applied to the kinematics.
In this case we consider a  purely leptonic final state (in this example $\mu^+\tau^-$), 
thus the label is $LL$.
The factorizable radiation for the `no-cut' case depends on the masses of the charged fermions
through collinear logarithms. 

In Fig.~\ref{fig:nocuts}, in addition to the combined factorizable/non-factorizable effect, 
$LL^{\sss{nf/fact}}$, 
we show separately the  initial-final, $LL^{\sss{if/fact}}$, and 
final-final, $LL^{\sss{ff/fact}}$, state ratios.
Note that they have opposite  signs.%
\footnote{This agrees with the observations of Ref.~\cite{kos} where  gluon radiation 
	in $e^+e^-\to t\bar{t}\to bW^+\bar{b}W^-$ was discussed.}
One can see from the figure that the final-final part of the non-factorizable correction
scales as $\alpha \sim E^{-4}$ with the CMS energy. In fact the power-counting arguments of 
Ref.~\cite{nf-anz} are applicable
to the parameter $\alpha$,  with a small modification due to the fixed rather than integrated
photon energy which does not however change the result. 
For the initial-final part of the interference the energy scaling is different:
$\alpha \sim  E^{-2}$. 
One can also see that initial-final state interference dominates the non-factorizable effects
at high energies,%
\footnote{This is again in agreement with observations of Ref.~\cite{kos}.}
and thus the complete non-factorizable radiation contribution also scales as $E^{-2}$.
If one does not apply any cuts,
the ratio of non-factorizable to factorizable contributions is small, below $3\%$. 
This is mainly due to the enhancement of the collinear logarithms in the 
factorizable part of the radiation. However if one keeps the photon well separated from 
the charged particles, 
and thus well away from the collinear regions, as in the  $LL_\gamma$  ratio in 
Fig.~\ref{fig:nocuts} then the collinear logarithms are suppressed and the 
ratio increases considerably. For example, in Fig.~\ref{fig:nocuts} 
$\alpha\approx 10-30\%$ for  the $LL_\gamma$ $>50\deg$ cut, depending on the collider energy.


\section{Enhancement mechanisms for different external states}

In Ref.~\cite{dkos} two processes were considered in detail: 
gluon radiation in $e^+e^- \to t\bar{t}$ and 
photon radiation in $\gamma\gamma\to W^+W^-$. 
Both cases were considered for collision energies close 
to threshold. The process we are interested in, (\ref{ee->ww->4f}), 
differs from the  studies of $\cite{dkos}$ in two respects. 

{}First, we consider
higher collision energies, which are experimentally  more relevant. Moreover 
the pole expansion,
which we use in our calculations, does not apply at threshold. On the other hand,
as we have already seen, at higher-energies non-factorizable effects become relatively
small and therefore the sensitivity to $\Gamma_W$ is less \cite{nf,nf-anz}. 
Without any cuts, the non-factorizable corrections, (\ref{mtrx}), to distributions 
inclusive with respect 
to angles  scale as $E^{-2}$ (initial-final state interference)
or $E^{-4}$ (final-final state interference)  relative to the Born cross section.
As a result of this scaling behaviour, at
 $\sqrt{s}=184\GeV$ the ratio of non-factorizable to factorizable contributions to the
 photon spectrum is $\OO(1\%)$ or smaller.
The main objective of our studies is to enhance the non-factorizable effects by applying 
{\it angular cuts}. 

A second difference with the study of Ref.~\cite{dkos} is that there the effects of initial
state radiation were not fully considered. In terms of Eq.~(\ref{mtrx}),  the analysis of 
Ref.~\cite{dkos} was concerned  with only one of the non-factorizable effects, from 
final-final state interference. 
As we have already noted, for $e^+e^- \to W^{+}W^{-} + \gamma$ the situation 
is more complicated,  with three interference contributions:
two initial-final and one final-final state.

In Ref.~\cite{dkos} two ways to enhance the non-factorizable effects were proposed.
First, it was suggested that certain angular asymmetry properties of 
initial-final state interference, absent in final-final state interference,
could be used  to construct observables to 
which initial-final state interference does not contribute.
Moreover, in \cite{dkos} an observable was constructed which has no contributions from 
factorizable radiation, 
 using the fact that the factorizable correction does not depend on the angles
 of the produced particles, 
at least at threshold (see Ref.~\cite{dkos} for more details).

Unfortunately, the construction of such observables is impractical. Because of 
$t$-channel neutrino exchange, there are always spin-charge correlations 
present, even in the threshold Born cross section.
Initial-final spin correlations induced by the $W$ propagators lead to
an asymmetry in the {\it factorizable} part of the radiation, as well as in the
 final-final 
state interference contributions. 
Therefore the method proposed in Ref.~\cite{dkos} does not appear to be workable.
Note that this effect originates in the 
$(v-a)$ structure of the charged weak current. The Born DPA cross section does not 
violate $P$-parity because only pole residues are calculated. 
However the type of helicity-charge correlation described above  does survive.
Technically, anti-symmetric tensors, $\epsilon_{\mu\nu\rho\sigma}$, induced by the 
axial current do not 
contribute linearly to the matrix element  (no $P$-violation), 
but only quadratically, via the  interference of
axial contributions from various stages of the process (the helicity-charge asymmetry).%
\footnote{Note that in $W$ pair production in $\gamma\gamma$ collisions the asymmetry in the 
threshold Born cross section 
	is absent, because there is no $(v-a)$ structure at the production stage. 
	Therefore, all the results of Ref.~\cite{dkos} remain valid for 
	the $\gamma\gamma\to W^+W^-$ case.
	The asymmetry is also absent in 
	$e^+e^- \to ZZ$ production, because of Bose symmetry. 
	On the other hand the asymmetry {\it will} be present,
	even without the  $(v-a)$  coupling of initial-state 
	fermions (as in QED for example),  if the initial-state fermions are polarized.}

Another idea discussed in \cite{dkos} was based on the  `angular ordering' effect.
During the last two decades such angular ordering effects have been intensively discussed
in the context of the QCD cascades, see for example \cite{ang-ordering}.
The phenomenon itself has been well known for QED in cosmic ray physics from 
the middle of the 1950s  as the so-called Chudakov effect \cite{chudakov}.
To recall the physics of angular ordering, we  consider the radiation pattern
of soft photons produced by a relativistic $e^+e^-$ pair. If we split the 
radiation into pieces associated with the $e^-$ and $e^+$, and then integrate over the azimuthal
angle about, say, the $e^-$ direction, the $e^-$ contribution vanishes for polar angles 
greater than the $e^+e^-$ opening angle. In particular this implies that the radiation 
vanishes for collinear $e^+$ and $e^-$. In other words, for such a configuration the 
emitted photon probes only the {\it total} electric charge of the $e^+e^-$ pair,
which is zero. The  suppression of radiation is caused by the (destructive) 
interference between the emission off the $e^-$
and $e^{+}$, see the second reference in \cite{dkos} for more details.
Because in the present context it is the  $W$ lifetime that 
controls this interference pattern, we expect to observe
angular ordering behaviour (or not)  according to the size of the ratio  $\omega/\Gamma_W$.
It is therefore  clear that the largest effect of non-factorizable 
radiation relative  to factorizable radiation will correspond to the case of 
collinear oppositely charged particles.
In that case factorizable radiation is as important as non-factorizable radiation.%
\footnote{
Note that due to the celebrated Low-Kroll-Barnett soft bremsstrahlung 
theorem \cite{low-theorem} the non-classical short-distance-induced 
corrections to the angular ordering behaviour arise only on the level of 
quadratic in $\omega/M$ terms, see Ref.~\cite{dks}.}

In the case of  $W$ pair production with $\mu\nu_\mu\tau\nu_\tau$ decay
in the  threshold region there are four radiating charged particles:
two initial state fermions, $e^{\pm}$, and two final state fermions, $\mu^+, \tau^-$. 
Corresponding to this there are three non-factorizable interferences: two initial-final, 
and one final-final state interference. Clearly  it is impossible to
generate  a large effect from all three interferences simultaneously. Indeed, if
the $e^+$ and $\mu^+$ are collinear and the $e^-$ and $\tau^-$ are collinear, then
the $\mu^+$ and $\tau^-$ are anti-collinear. 
Far above threshold, the  directions of the $W$-boson momenta start to play a role as well.
Thus in general one has many cases when some oppositely charged particles are collinear
and others are not, leading to a non-trivial interplay between 
the  various interference terms in (\ref{mtrx}). Rather than choose particular fixed 
configurations, for which the statistics will be small, it is more efficient 
to look for angular cuts (no-flight zones) on 
the various particles such that the interesting (i.e.
most $\Gamma_W$ dependent) events are favoured but not overly restricted.
We shall not in the present study make any serious attempt to {\it optimize} these cuts;
rather we will present some illustrative examples pending more detailed Monte Carlo
analyses.

In summary,  angular cuts (no-flight zones)
will be applied to the final state particles (leptons and quarks)
and the photon in order to maximize the 
angular ordering
effect, and thus the sensitivity of the photon spectrum to the $W$ width.
As explained above, the basic idea is to 
keep oppositely charged particles quasi-collinear,
and the photon as far from them as possible. 
There is an additional requirement motivated by detector considerations.
The  final state particles 
should not be too close to the beam direction otherwise the event cannot be 
unambiguously identified as $W^+W^-\gamma$.
We therefore require all final state particles to be produced at polar
angles greater than $5\deg$ from the beam direction.%
\footnote{We are grateful to G.~Wilson for clarification of various experimental issues
	related to $W$ studies at a future linear collider.}

\subsection{Leptonic-leptonic final state}

The simplest case to analyse is when both $W$'s
decay leptonically:
\be
\label{ee->ll}
        e^{+}(q_1)\; +\; e^{-}(q_2)\;	 \to\;	 \mu^{+}(p_3)\; + \; \tau^-(p_4)\; +
        \; 2\nu	\; + \; \gamma(k)\; .
\ee
The first topology we will consider is when the two final-state charged particles are 
close to the beam direction. In this case the 
initial-final state interference gives a large effect. 
In addition, the photon should be far from the beam direction:
\be
                \angle(q_{1,2} k) > 50\deg \; .
\ee
The charged final-state leptons with momenta $p_3$ and $p_4$ can each be
 either collinear to 
the initial-state positron or to the electron. `Collinear' is here defined
as being produced with polar angle between $5\deg$ and $10\deg$ 
with respect to the beam direction: 
\be
	 \angle(q_1 p_{3,4}) \in (5\deg,10\deg)\; , \ \ \ \mbox{or} \ \ \ 
         \angle(q_2 p_{3,4}) \in (5\deg,10\deg) \; .
\ee
We therefore have four possible cases, which we label
\be
	LL_{++}\; , \ \ \ LL_{+-}\; , \ \ \ LL_{-+}\; , \ \ \ LL_{\times\times} \; ,
\ee
corresponding to  
($p_3\parallel q_1$ and $p_4\parallel q_1$),
($p_3\parallel q_1$ and $p_4\parallel q_2$), and
($p_3\parallel q_2$ and $p_4\parallel q_1$) correspondingly.
$LL$ refers to the fact that both $W$ bosons decay leptonically.
In the last case $LL_{\times\times}$ we demand only that the final-state 
leptons are collinear with the  electron and positron beams, 
without tracing the electric charge flow. 

The second class of cuts
we will consider is when two final-state particles are quasi-collinear. 
In this case it is the final-final state interference that produces a large effect.
We first demand that all final-state particles are observable
\be
	\angle(q_{1,2} p_{3,4}) > 5\deg\; ,
\ee
and then that the final-state charged particles are collinear
\be
	\angle(p_3 p_4) < 10\deg\; ,
\ee
and the photon is far from all charged particle directions
\be
	\angle(p_{3,4} k) > 50\deg \; , \ \ \ 
	\angle(q_{1,2} k) > 50\deg \; .
\ee
Here there is only  one possible case, which we label
\be
	LL_{\sss{f}}\; .
\ee
\begin{figure}[t]
  \unitlength 1cm
  \begin{center}
  \begin{picture}(8,7)
  \put(-0.2,6){\makebox[0pt][c]{\boldmath\small $\alpha$}}
  \put(6.5,-0.3){\makebox[0pt][c]{\boldmath\small$\sqrt{s}, GeV$}}
  \put(0.2,0){\epsfig{file=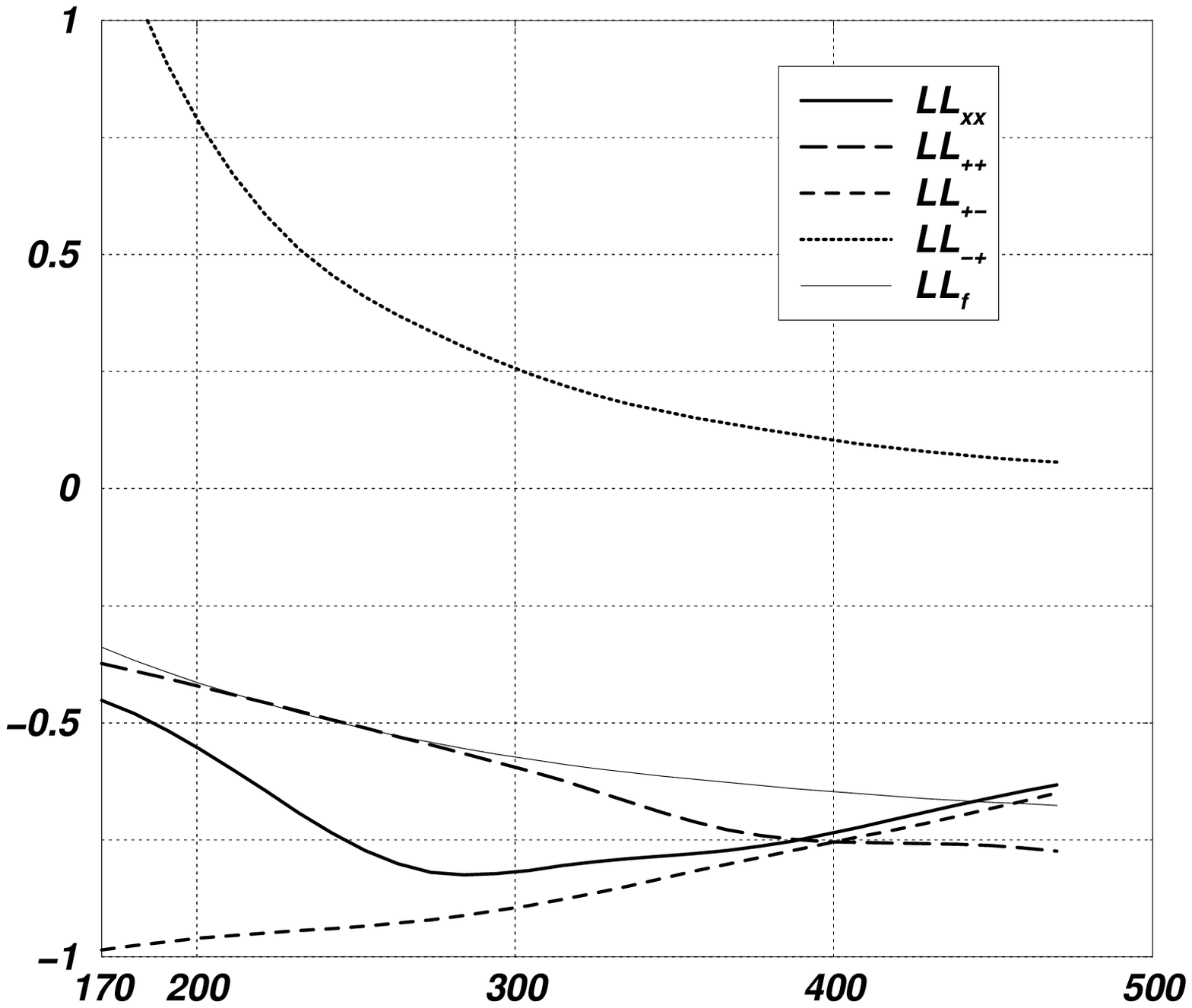,height=7cm,width=7cm,angle=0}} 
  \end{picture}
 \begin{picture}(8,7)
  \put(-0.2,6){\makebox[0pt][c]{\boldmath\small $\sigma_{0}$}}
  \put(-0.2,5){\makebox[0pt][c]{\boldmath\small [pb]}}
  \put(6.5,-0.3){\makebox[0pt][c]{\boldmath\small$\sqrt{s}, GeV$}}
  \put(0.2,0){\epsfig{file=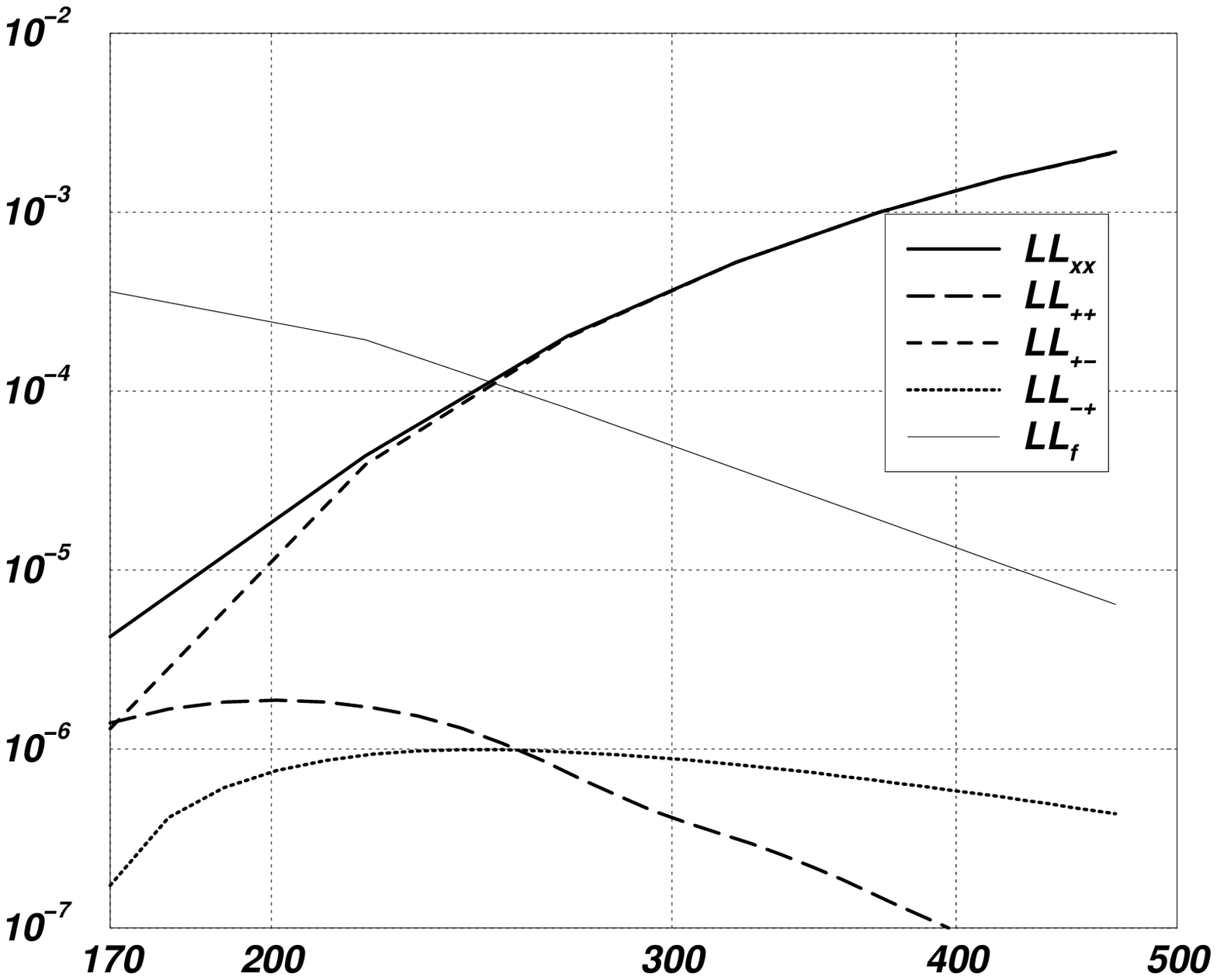,height=7cm,width=7cm,angle=0}}
  \end{picture}
  \end{center}
  \caption[]{Ratio of the non-factorizable and factorizable parts of the photon
                 radiation, $\alpha$, 
		and the Born cross section, $\sigma_{\sss{Born}}$, as a function of the
		CMS energy and the 
		system of cuts. Purely leptonic final state only.}
\label{fig:LL}
\end{figure}

The optimization parameters $\alpha(\sqrt{s},\mbox{cuts})$ and $\sigma_{\sss{Born}}(\sqrt{s}, \mbox{cuts})$
are shown in Fig.~\ref{fig:LL} as functions of the CMS energy, $\sqrt{s}$, 
for all five possible leptonic cuts.
We see that at low energies the most pronounced shape of the photon spectrum is achieved
for the $LL_{-+}$ case, i.e. $\mu^+$ collinear with incoming $e^-$ and
$\tau^-$ collinear with incoming $e^+$. Then the ratio of the  non-factorizable to 
and factorizable 
contributions is positive and can even exceed 1, in the lower energy domain. 
However, the Born cross-section for this set of cuts is very small, 
which makes this case
statistically disadvantageous. At high energies, the outgoing fermion (antifermion)
 prefers to
follow the direction of the incoming fermion (antifermion), and hence both the
$LL_{+-}$  and  $LL_{\times\times}$ configurations have large Born cross sections, 
$\sigma_{\sss{Born}}$.
In terms of the shape parameter, the $LL_{+-}$ cut is as good as $LL_{\times\times}$.
At lower energies, however, $LL_{+-}$  becomes more advantageous in terms of shape, 
but less advantageous in terms of statistics. In fact, referring back to
Fig.~\ref{fig:nocuts}, we see that the original
$LL_\gamma$-cut is as good in terms of shape
as $LL_{\times\times}$ at low energy, but much better statistically since it corresponds to
a much larger angular acceptance for the final-state charged particles.
At higher energies it is still as good in terms of shape, but becomes 
 statistically very poor.

  The conclusion is that 
depending on the energy and statistics available, one can choose different 
systems of cuts as the preferred ones. There is clearly
 no unique `best cut' for all energies and all statistics.

\subsection{Leptonic-hadronic final state}

We next consider the case when the $W^{+}$ decays leptonically and 
the $W^{-}$ decays hadronically.
There is one charged lepton and two jets present in the final state:
\be
\label{ee->lqq}
        e^{+}(q_1)\; +\; e^{-}(q_2)\;	 \to\;	 \mu^{+}(p_3)\; + \; q(p_4)\; +
        \;  \bar{q}'(p_4')
        \; + \nu	\; + \; \gamma(k)\; .
\ee
We again start with the case when the charged primary fermions are close to the 
beam directions.
Just as in the lepton-lepton case,  we demand that the photon is far from the beams,   
$\angle(q_{1,2} k) > 50\deg$ and the final-state lepton with momenta $p_3$ 
is either collinear to  the initial-state positron or electron,
$\angle(q_1 p_3) \in (5\deg,10\deg)$ or $\angle(q_2 p_3) \in (5\deg,10\deg)$.
The two quarks with momenta $p_4$  and $p_4'$ should also be collinear to the 
initial-state particles:
\be
	 \angle(q_1 p_4) \in (5\deg,20\deg)\; , \ \ \ \mbox{or} \ \ \ 
         \angle(q_2 p_4) \in (5\deg,20\deg)\; , 
\ee
and
\be
	\angle(q_1 p_4') \in (5\deg,20\deg)\; , \ \ \ \mbox{or} \ \ \ 
         \angle(q_2 p_4') \in (5\deg,20\deg)\; .
\ee
An important difference here is that one cannot measure the charge of the jet 
experimentally. Thus in general the following combinations are available:
$$
	LH_{+(20)}\; , \ \ \ 
	LH_{+(02)}\; , \ \ \ 
	LH_{-(20)}\; , \ \ \ 
	LH_{-(02)}\; , \ \ \ 
$$
$$	
	LH_{+(11)}\; , \ \ \ 
	LH_{-(11)}\; , \ \ \ 
	LH_{+(\times\times)}\; ,  \ \ \ 
	LH_{-(\times\times)\; }, \ \ \ 
	LH_{\times(\times\times)}\; ,
$$
where $LH$ denotes the  leptonic-hadronic final state.
The first subscript indicates the direction of the lepton with respect 
to the positron momentum,  and the two subscripts in parenthesis
indicate the number of jets collinear with the positron and electron. 
For example, $LH_{+(20)}$ means that the 
final-state lepton is collinear with the positron, as are both jets.
$LH_{-(\times\times)}$ means that the  final-state lepton is collinear with the
 electron, and the two jets are 
collinear to either the positron or electron. Thus, in general, the
number of different cases is rather large compared to the purely leptonic final state.
For the energies we are interested in, $\sqrt{s}=170-500\GeV$, however, 
the situation simplifies somewhat, because the kinematics are such that
the two jets coming from the
 decay of the $W$ boson
cannot in fact satisfy the collinearity selection criterion. 
Thus
$$
	LH_{+(20)}=
	LH_{+(02)}=
	LH_{-(20)}=
	LH_{-(02)}=0 \; .
$$	
The following cases survive
\be
        LH_{+(11)} = LH_{+(\times\times)} \equiv LH_{+\times}\; , \ \ \ 
        LH_{-(11)} = LH_{-(\times\times)} \equiv LH_{-\times}\; , \ \ \ 	
	LH_{\times(\times\times)}\equiv LH_{\times\times}\; ,
\ee
where we have introduced the modified 
notation $LH_{+\times}$, $LH_{-\times}$ and $LH_{\times\times}$.

The second class of cuts  again corresponds to the situation when
two final-state particles are  collinear. 
In this case the final-final state interference gives a large effect.
We demand that all final-state particles are observable
\be
	\angle(q_{1,2} p_{3,4}) > 5\deg\; , \ \ \ 	
	\angle(q_{1,2} p_4') > 5\deg\; ,
\ee
at least two final-state particles are quasi-collinear
\be
	\angle(p_3 p_4) < 10\deg\; , \ \ \ \mbox{or} \ \ \ 
	\angle(p_3 p_4') < 10\deg\; ,
\ee
and the photon is far from all charged particles
\be
	\angle(p_3 k) > 50\deg\; ,	\ \ \ 
	\angle(p_4 k) > 50\deg\; ,	\ \ \  
	\angle(p_4' k) > 50\deg\; ,	\ \ \  	
	\angle(q_{1,2} k) > 50\deg\; .
\ee
Thus there is again only one possible choice
\be
	LH_{\sss{f}}\; .
\ee
\begin{figure}[t]
  \unitlength 1cm
  \begin{center}
  \begin{picture}(8,7)
  \put(-0.2,6){\makebox[0pt][c]{\boldmath\small $\alpha$}}
  \put(6.5,-0.3){\makebox[0pt][c]{\boldmath\small$\sqrt{s}, GeV$}}
  \put(0.2,0){\epsfig{file=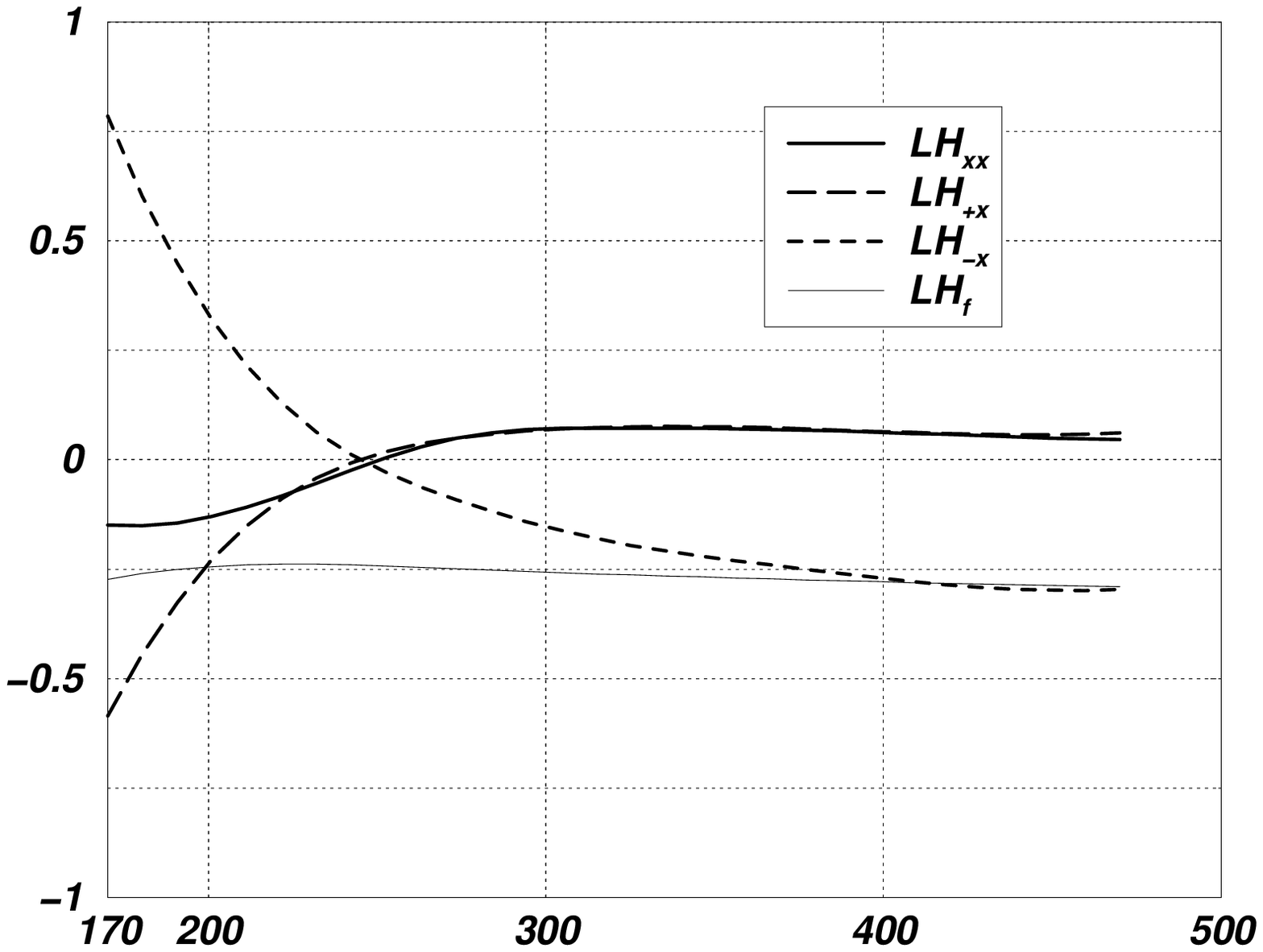,height=7cm,width=7cm,angle=0}} 
  \end{picture}
 \begin{picture}(8,7)
  \put(-0.2,6){\makebox[0pt][c]{\boldmath\small $\sigma_{0}$}}
  \put(-0.2,5){\makebox[0pt][c]{\boldmath\small [pb]}}
  \put(6.5,-0.3){\makebox[0pt][c]{\boldmath\small$\sqrt{s}, GeV$}}
  \put(0.2,0){\epsfig{file=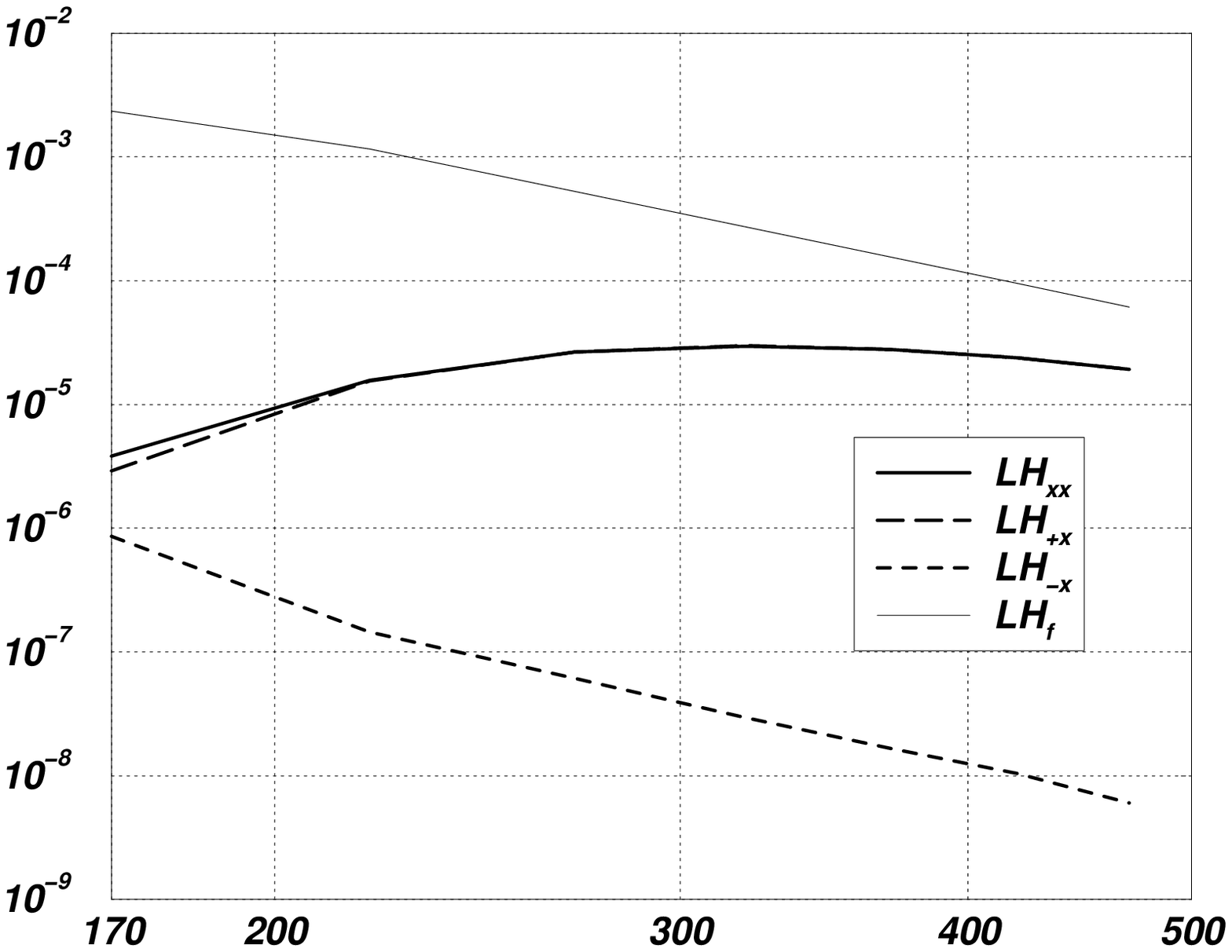,height=7cm,width=7cm,angle=0}}
  \end{picture}
  \end{center}
  \caption[]{Ratio of the non-factorizable and factorizable parts of 
  the photon radiation, 
		$\alpha$, 
		and the Born cross section, $\sigma_{\sss{Born}}$, as a function of the 
		CMS energy and the 
		system of cuts. Leptonic-hadronic final states.}
\label{fig:LH}
\end{figure}%

We show in Fig.~\ref{fig:LH} the optimization parameters 
$\alpha(\sqrt{s},\mbox{cuts})$ and 
$\sigma_{\sss{Born}}(\sqrt{s}, \mbox{cuts})$
 as functions of the CMS energy, 
$\sqrt{s}$, for all possible $LH$-cuts. Again, we see that the $LL_{-\times}$-cut is the 
best in terms of the shape of the spectrum, but at the same time it is the worst in terms of 
statistics. $LH_f$ seems to be statistically the best overall throughout the energy region
under consideration. From the point of view of the 
shape of the spectrum, $LH_f$ is not worse than 
any other system of cuts for higher energies. Note that 
for cuts restricting the jets to be quasi-collinear with the collider beams,
the energy behaviour of the shape parameter $\alpha$
is more complicated than in the case of leptonic-leptonic final states. 
As  mentioned above, if one of the quarks is quasi-collinear with the 
 electron, the other one is automatically quasi-collinear with the positron. 
 The effects coming from the two corresponding interferences have opposite signs.
This can even lead to a change of sign of the combined effect at 
different collider energies.

\subsection{Hadronic-hadronic final state}

Finally we consider the case when both $W$ bosons decay hadronically, with
four jets present in the final state.

We again start from the case when the charged particles (i.e. jets) are collinear with the beam direction,
with the photon well separated from the beam, $\angle(q_{1,2} k) > 50\deg$.
The quarks with momenta $p_{3,4}$  and $p_{3,4}'$ are required to be collinear with the initial leptons,
$\angle(q_{1,2} p_{3,4}) \in (5\deg,20\deg)$ and $\angle(q_{1,2} p_{3,4}') \in (5\deg,20\deg)$.
Since  one cannot measure the charge of the jet experimentally.
only the following combinations are available:
$$
	HH_{22}\; , \ \ \ 
	HH_{13}\; , \ \ \ 
	HH_{\times\times}\; ,
$$
where the subscript denotes the  number of jets that are collinear with the initial-state positron or electron.
For example, $HH_{13}$ means that there is one jet collinear with the positron, $q_1$,
 and three jets collinear
with the electron, $q_2$. Again, the kinematics are such that at LEP2 energies 
not all of these cases are non-zero:
$$
	HH_{13}=0\; ,
$$
and in fact only one case survives:
\be
	HH_{22} = HH_{\times\times}\; .
\ee

The second class of cuts corresponds to  when two final-state particles (jets) are quasi-collinear. 
In this case final-final state interference gives a large effect.
We  first demand that all final state particles are observable
\be
	\angle(q_{1,2} p_{3,4}) > 5\deg\; , \ \ \ 	
	\angle(q_{1,2} p_{3,4}') > 5\deg\; ,
\ee
at least two final state particles are collinear
\be
	\angle(p_3 p_4) < 10\deg\; , \ \ \ \mbox{or} \ \ \ 
	\angle(p_3 p_4') < 10\deg\; , \ \ \ \mbox{or} \ \ \ 
	\angle(p_3' p_4) < 10\deg\; , \ \ \ \mbox{or} \ \ \ 
	\angle(p_3' p_4') < 10\deg\; ,
\ee
and the photon is far from all of the charged particles
\be
	\angle(p_{3,4} k) > 50\deg\; , \ \ \ 
	\angle(p_{3,4}' k) > 50\deg\; ,	\ \ \  	
	\angle(q_{1,2} k) > 50\deg\; .
\ee
There is again only one possible case:
\be
	HH_{\sss{f}}\; .
\ee
\begin{figure}[t]
  \unitlength 1cm
  \begin{center}
  \begin{picture}(8,7)
  \put(-0.2,6){\makebox[0pt][c]{\boldmath\small $\alpha$}}
  \put(6.5,-0.3){\makebox[0pt][c]{\boldmath\small$\sqrt{s}, GeV$}}
  \put(0.2,0){\epsfig{file=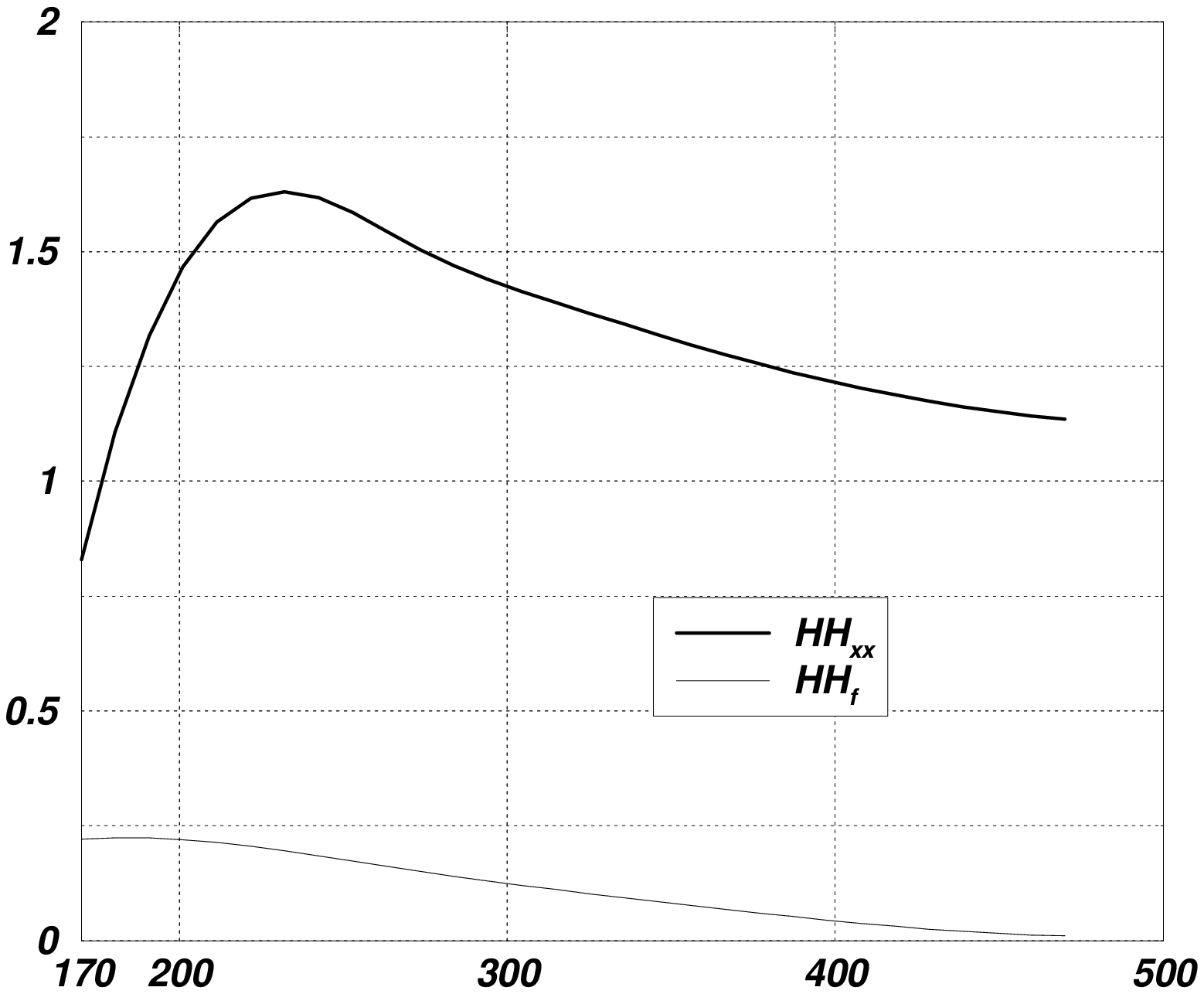,height=7cm,width=7cm,angle=0}} 
  \end{picture}
 \begin{picture}(8,7)
 \put(-0.2,6){\makebox[0pt][c]{\boldmath\small $\sigma_{0}$}}
  \put(-0.2,5){\makebox[0pt][c]{\boldmath\small [pb]}}
  \put(6.5,-0.3){\makebox[0pt][c]{\boldmath\small$\sqrt{s}, GeV$}}
  \put(0.2,0){\epsfig{file=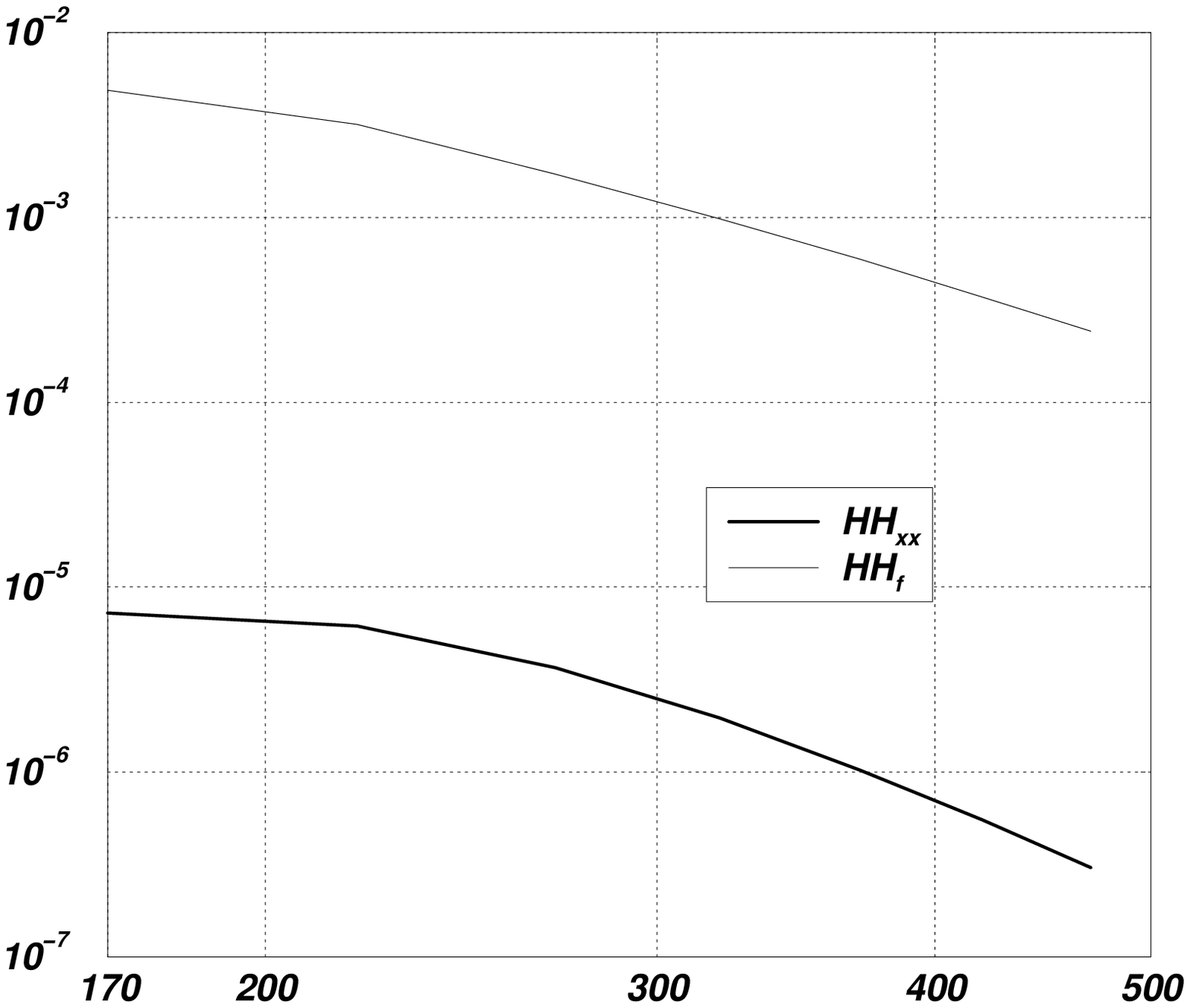,height=7cm,width=7cm,angle=0}}
  \end{picture}
  \end{center}
  \caption[]{Ratio of the non-factorizable and factorizable parts of the photon radiation, 
		$\alpha$, 
		and the Born cross section, $\sigma_{\sss{Born}}$, as a function of the
		CMS energy and 
		the system of cuts. Hadronic-hadronic final states.}
\label{fig:HH}
\end{figure}%

The optimization parameters $\alpha(\sqrt{s},\mbox{cuts})$ and 
$\sigma_{\sss{Born}}(\sqrt{s}, \mbox{cuts})$ are shown in Fig.~\ref{fig:HH}
for the  two possible cuts $HH_{\times\times}$ and $HH_{f}$.
As in the previous cases, $HH_{f}$ cut is better statistically, but $HH_{\times\times}$
is better from the point of view of the shape of the photon spectrum.

\subsection{Photon-photon colliders}

In recent  years there has been a growing interest in high-energy photon colliders, using 
Compton back-scattering of laser light off the lepton beams at
linear colliders to produce high-intensity, high-energy beams of photons, see e.g. Ref.~\cite{telnov}.
Using $\gamma\gamma$ collisions to produce pairs of $W$ bosons offers
certain advantages over the $e^+e^-$ case. First, the cross section is an order of magnitude larger.
Second,  ISR effects are absent in this case and so kinematic reconstruction
of the $WW$ final state is in principle more precise.
 
It is straightforward to extract the predictions for photon radiation in $\gamma\gamma\to W^+W^-$
{}from our study of the more complicated $e^+e^-$ case.
In particular,  our results for the final-final state interferences $LL_{\sss{f}}$,
$LH_{\sss{f}}$ and $HH_{\sss{f}}$ can be applied directly to  the $\gamma\gamma$ case.
Moreover, as we have already explained, in the case of $W^+W^-$ production
in photon-photon  collisions one can study observables integrated over the photon angle, to which 
factorizable corrections do not contribute, see Ref.~\cite{dkos}.%
\footnote{Note that there is a typo in Eq.~(20) of the first reference in \cite{dkos}.
	The normalization coefficients in front of the integrals in the first and second terms
	should be interchanged. The final result given by Eq.~(22) is unchanged.}
This enables us to utilize more events and makes studies in $\gamma\gamma$ collisions
potentially more statistically powerful than in the $e^+e^-$-case.


\section{Concluding remarks}

A precision measurement of the total $W$ decay width presents a challenge for present and future
experiments. Line-shape measurements are made difficult by the presence of neutrinos in the final state
in the case of leptonic decay modes, and of hadronization corrections in the case of hadronic decays.
The indirect measurement at hadron colliders, which uses the ratio of $W$ and $Z$ leptonic
events, has an inherent uncertainty from parton distributions in the theoretical calculation 
of the total cross sections.
It seems to be quite a challenging task to perform a precise direct measurement of the total $W$-width,
independent of decay modes (and of the $Z$ measurements). 

As discussed in Ref.~\cite{wilson},  running a future linear collider
in the `LEP2' energy region may provide a unique opportunity for a high-precision
measurement of the $W$ mass and width.
The 'traditional' way of measuring $\Gamma_W$  is from  a threshold scan of the total
$WW$ cross section.  Though  
statistically powerful, this method is not without problems. The uncertainties caused by
beam-induced effects (beamsstrahlung, intrinsic energy spread, etc.) could be potentially large.
Moreover, the threshold strategy requires operating a linear collider at energy 
scan points well below threshold where the 
$W^+W^-$ cross section is very small. 

In this paper we have argued that the soft-photon radiation spectrum could also
be used to obtain information on $\Gamma_W$. We emphasize that this is an independent method ---
in effect one is measuring the non-factorizable interference  to the cross section,
whose magnitude is controlled by the relative size of the photon energy  and the $W$ width.
The method is in principle very clean, requiring only a precise measurement of
the soft (i.e. of order few GeV) photon spectrum in $W^+W^-\gamma$ events. However, as we have seen,
the effect in the inclusive distribution is very small and therefore is likely to be limited
by statistics. On the other hand, we have shown that one can enhance the effect
by employing angular cuts on the final-state particles. We have considered various different
topologies and  different $W$ decay channels. Both the sensitivity  to the non-factorizable contributions
and the overall number of events in the various channels are rather strongly dependent
on the collision energy, and it should be possible to develop an optimal strategy given
the parameters and running conditions of a future linear collider.

Our study necessarily falls short of any firm conclusion about the competitivity of our method,
compared to the threshold scan for example, in determining $\Gamma_W$.
At the very least, our method offers a complementary measurement, with completely different
systematics. The next step would be to perform a detailed Monte Carlo study including 
detector and, where appropriate, hadronization effects. Among the questions that
such a study could answer are: what is the efficiency for detecting very soft photons?
can such photons be measured in the presence of hadronic jets? are the isolation and collinearity
cuts we have used realistic? For a given collider energy it should be straightforward
to estimate the number of soft photon events for each of the different topologies and decay channels,
and by comparing this with the theoretical predictions, to estimate the statistical error
on $\Gamma_W$. The results of our work suggest that a more detailed study is definitely
worth pursuing.

\vskip 1cm

\noindent{\bf Acknowledgements}\\
We thank T.~Sj\"ostrand, L.~Stodolsky and G.~Wilson for useful discussions. 
VAK thanks the Leverhulme Trust for a Fellowship. The work of APC is supported in part
by the UK Particle Physics and Astronomy Research Council.
This work was also supported by
EU Fourth Framework Programme ``Training and Mobility of Researchers'',
contract FMRX-CT98-0194 (DG 12 - MIHT).


\end{document}